# Effect of processing conditions on the thermal and electrical conductivity of poly (butylene terephthalate) nanocomposites prepared via ring-opening polymerization

S. Colonna[a], M.M. Bernal[a], G. Gavoci[a], J. Gomez[b], C. Novara[c], G. Saracco[d], A. Fina[a,*]

[a]Dipartimento di Scienza Applicata e Tecnologia, Politecnico di Torino, 15121 Alessandria, Italy
[b]AVANZARE Innovacion Tecnologica S.L., 26370 Navarrete (La Rioja), Spain
[c]Dipartimento di Scienza Applicata e Tecnologia, Politecnico di Torino, 10129 Torino, Italy
[d]Istituto Italiano di Tecnologia, Centre for Sustainable Futures CSF@PoliTo, 10129 Torino, Italy
*Corresponding author: alberto.fina@polito.it

**Abstract**

Successful preparation of polymer nanocomposites, exploiting graphene-related materials, via melt mixing technology requires precise design, optimization and control of processing. In the present work, the effect of different processing parameters during the preparation of poly (butylene terephthalate) nanocomposites, through ring-opening polymerization of cyclic butylene terephthalate in presence of graphite nanoplatelets (GNP), was thoroughly addressed. Processing temperature (240°C or 260°C), extrusion time (5 or 10 minutes) and shear rate (50 or 100 rpm) were varied by means of a full factorial design of experiment approach, leading to the preparation of polybutylene terephthalate/GNP nanocomposite in 8 different processing conditions. Morphology and quality of GNP were investigated by means of electron microscopy, X-ray photoelectron spectroscopy, thermogravimetry and Raman spectroscopy. Molecular weight of the polymer matrix in nanocomposites and nanoflake dispersion were experimentally determined as a function of the different processing conditions. The effect of transformation parameters on electrical and thermal properties was studied by means of electrical and thermal conductivity measurement. Heat and charge transport performance evidenced a clear correlation with the dispersion and fragmentation of the GNP nanoflakes; in particular, gentle processing conditions (low shear rate, short mixing time) turned out to be the most favourable condition to obtain high conductivity values.

**Keywords:** conductive polymer nanocomposites; graphite nanoplatelets; reactive extrusion; thermal conductivity; nanocomposites

**1. Introduction**

The growing demand for the substitution of metals in those applications where heat exchange coupled with corrosion resistance, ease of process and low cost are required, pushed research on polymer composites and nanocomposites [1-5]. Here, the addition of proper fillers leads to the obtainment of thermally conductive materials, despite pristine polymers behave as heat insulators. Fillers to be used for this application include graphite, carbon fibers (CF), carbon nanotubes (CNT), graphene-related materials (GRM), hexagonal boron nitride (hBN), metal and ceramic powders, etc. and were extensively studied in literature [6]. As carbonaceous fillers are both thermally and electrically conductive, their polymers nanocomposites may also be used as functional materials for applications such as sensors and actuators [7-9], including shape memory polymers [10-12].

The discovery of graphene in 2004 [13], defined as a single-atom-thick sheet of hexagonally arranged $sp^2$-bonded carbon atoms, and its outstanding thermal, mechanical and electrical properties [14-16] attracted a lot of interest in the scientific community. Despite the development of different synthetic techniques [17-19], the manufacturing of high quality graphene, *i.e.* low defective flakes with high lateral size, is still associated to the production of very limited quantities of material, whereas exploitation of graphene in polymers for industrial applications requires large scale production. In fact, nanoparticles available in sufficient amount to be exploited in polymer composites are typically from chemical reduction of graphene oxide (GO) [20], ball milling [21] and thermal exfoliation and reduction of GO [22]. However, it is worth noting that these large scale production processes typically lead to the synthesis of so called graphene-related materials rather than single layer graphene, including graphite nanoplatelets (GNP), reduced graphene oxide (rGO) multi-layer graphene (MLG), etc. [23], with various degree of structural and chemical defectiveness.

The improvement of thermal conductivity in polymer/GRM nanocomposites is mainly related to the quality of carbon nanoflakes, their organization in the polymer matrix and the interfacial thermal conductances [24]. The importance of flakes quality was demonstrated by molecular dynamic simulation [25] and recently verified experimentally [26] with a dramatic decreases of the intrinsic thermal conductivity of graphene as a function of defect concentration; furthermore, we reported that reduction of defects upon annealing at 1700°C in vacuum, raises the thermal conductivity of rGO [27]. Additionally, it was demonstrated in literature that higher lateral size is straightforwardly related to a higher thermal conductivity [28, 29]. In a recent paper, we demonstrated that the addition of high-temperature-annealed rGO in a polymer matrix leads to a thermal conductivity which is about 2-fold those of poly (butylene terephthalate) containing pristine rGO (higher defectiveness) or GNP [30], further confirming the need of high quality nanoflakes for the preparation of highly thermally conductive polymer nanocomposites. On the other hand, the control of GRM organization into a polymer matrix remains crucial in terms of nanoparticle distribution and quality of contacts between particles. Attempts in precisely controlling orientation and contacts between nanoparticles indeed resulted in an improvement of thermal transfer [31-33] but the methods adopted for the preparation of these nanocomposites are hardly up-scalable or requires very high filler concentrations. Finally, the reduction of interfacial thermal resistance was also pursued by nanoparticle functionalization [34-36], despite the effectiveness of this strategy may be also related to the lateral size of the nanoflakes [37]. Recently, the preparation of polymer nanocomposites was obtained by in-situ ring-opening polymerization of cyclic butylene terephthalate (CBT) oligomers into poly (butylene terephthalate), pCBT [38], taking advantage of both the extremely low viscosity of CBT and of the viscosity increase occuring during polymerization, to disperse nanoparticles, including organoclays [39], carbon nanotubes [40], silica [41] and graphene-related materials [30, 42]. The presence of GRM was reported to affect the polymerization kinetic of CBT, with increase of the polymerization time [43] and decrease of the average molecular weight [44]. Furthermore, the exploitation of GRM was described to improve mechanical, electrical and thermal properties [30, 42, 43, 45]. Indeed, in a previous paper we reported a 12-fold increase in the thermal conductivity when 30 wt.% of GNP is added to pCBT, while the addition of 5 wt.% of rGO annealed at 1700°C led up to a 4-fold increase [30].

In this work, the optimization of different processing paremeters in the preparation of pCBT nanocomposites through ring-opening polymerization of CBT is addressed. In particular, the effect of processing temperature, mixing time and shear rate on electrical and thermal conductivity of nanocomposites are described in this work, aiming at a systematic study of processing conditions *vs.* material properties, which is still lacking in the field of polymer nanocomposites containing graphene-related materials

## 2. Experimental

### 2.1. Materials

Cyclic butylene terephthalate oligomers [CBT100, Mw = $(220)_n$ g/mol, $n$ = 2-7, melting point= 130 ÷ 160°C] were purchased from IQ-Holding[1] (Germany). Butyltin chloride dihydroxide catalyst (96%, $m_p$ = 150°C), Chloroform ($CHCl_3$) (≥ 99.9%) and 1,2 Dichlorobenzene (≥ 99%) were purchased from Sigma-Aldrich; Acetone (99+%), 1,1,1,3,3,3 Hexafluoroisopropanol (HFIP, ≥ 99%) and Phenol (≥ 99.5%) were purchased from Alfa Aesar, Fluka and Riedel-de Haën, respectively.

### 2.2. Synthesis of graphite nanoplatelets

The GNP used in this work was a research grade (see below for preparation method) synthetized by AVANZARE (Navarrete, La Rioja, Spain) using a rapid thermal expansion of overoxidized-intercalated graphite (ox-GIC). The intercalation of graphite with sulphuric acid to obtain graphite-sulphate is a well-known technology described for the first time by Hofmann and Rüdorff in 1938 [46]. In the present paper, the synthesis of GIC was made by adding 40 g of natural graphite flakes (average lateral size ≈ 1 mm) and 400 g of sulfuric acid ($H_2SO_4$) in a 5 liters refrigerated glass jacket reactor under continuous stirring at T < 10°C. Then, 5 g of nitric acid ($HNO_3$) were added drop by drop with a peristaltic pump, keeping the temperature constant. Later, 12.5 g of potassium permanganate ($KMnO_4$) were added to the suspension, keeping the temperature below 10°C. When $KMnO_4$ was completely added, the system was heated up to 50°C and stirred at this temperature for 1 hour to allow the completion of the reaction (indicated by a change on the color of the suspension, from brown to black). At this point the system was cooled to room temperature and the solution was pumped, with a peristaltic pump, into a tank of $H_2O$ (≈ 2 L), keeping the temperature lower than 70°C. Hydrogen peroxide ($H_2O_2$, 30 g, 30 v.%) was slowly added to remove the excess of $MnO_4^-$, and the suspension was maintained under stirring for about 30 min at room temperature. The solution was washed in 3 L of 3.3 wt.% HCl solution for 1h. Then, the solid was filtered, rinsed with osmotic water (until the sulfate test gave a negative result), dried in air and then in an oven at 80°C. The resultant black powder was mechanically milled in a ball mill. The obtained solid, named ox-GIC-1, was then introduced in a tubular furnace under inert atmosphere ($N_2$) at 1000°C for thermal expansion, obtaining a worm-like solid; this was later mechanically milled, separating nanoflakes and obtaining GNP.

## 2.3. Nanocomposite preparation

Polymer nanocomposite preparation consisted in a two-step procedure. In a first step, CBT/GNP mixture was prepared by mixing about 5 wt.% of GNP (with respect to CBT content) into an acetone/CBT solution (~ 0.15 g/mL). After 2 hours, the solvent was evaporated and the CBT/GNP mixture was dried at 80°C under vacuum (~$10^1$ mbar) to completely remove residual acetone and moisture. In a second step, the dried mixture was manually pulverized and loaded into a co-rotating twin screw micro-extruder (DSM Xplore 15, Netherlands) and mixed for ~ 5 minutes; then, 0.5 wt.% of butyltin chloride dihydroxide catalyst (with respect to the oligomer amount) was added and the extrusion proceeded to complete CBT polymerization into pCBT. To avoid thermo-oxidative degradation and hydrolysis of the organic matrix, the extrusion process was performed under an inert atmosphere.

Different processing temperature, extrusion time and shear rate were selected by means of a full factorial design of experiment (DOE) approach based on three parameters, for each of those defining a low and high value, leading to the preparation of 8 nanocomposites. Furthermore, pure pCBT was synthetized (extrusion parameters: 240°C, 5 minutes and 50 rpm) as reference material. The different nanocomposites are labeled as pCBT_GNP/*x*/*y*/*z* where *x* is the temperature in °C, *y* the extrusion time in min and *z* the shear rate in rpm. Processing conditions, coupled with nanocomposites labeling, are reported in Table 1.

**Table 1. Labeling and processing conditions used for pCBT and pCBT nanocomposites**

|   | Material | Processing parameters | | |
|---|---|---|---|---|
|   |   | Temperature [°C] | Time [Minutes] | Shear rate [rpm] |
| 1 | **pCBT** | 240 | 5 | 50 |
| 2 | **pCBT_GNP/240/5/50** | 240 | 5 | 50 |
| 3 | **pCBT_GNP/240/10/50** | 240 | 10 | 50 |
| 4 | **pCBT_GNP/240/5/100** | 240 | 5 | 100 |
| 5 | **pCBT_GNP/240/10/100** | 240 | 10 | 100 |
| 6 | **pCBT_GNP/260/5/50** | 260 | 5 | 50 |
| 7 | **pCBT_GNP/260/10/50** | 260 | 10 | 50 |
| 8 | **pCBT_GNP/260/5/100** | 260 | 5 | 100 |
| 9 | **pCBT_GNP/260/10/100** | 260 | 10 | 100 |

## 2.4. Intrinsic viscosity determination

The nanocomposites were dissolved in a mixture solvent of $CHCl_3$/HFIP (90/10 v/v) for ~ 1 hour at room temperature, and filtered through a PTFE membrane (0.45 μm pore size) to separate GNP (efficiency of polymer extraction ~ 98 %, calculated by TGA). The polymer solution was concentrated under reduced pressure and dried at 80 °C overnight.

Intrinsic viscosity measurements [η] were performed with a Type II Ubbelohde capillary viscometer at 25 °C in a mixture of phenol/1,2-dichlorobenzene (50/50 w/w), according to the ISO 1628-5. The pCBT samples were dissolved in the above mixture at 75 °C until complete solution was achieved (~ 1 hour). The solution was then cooled to room temperature and the intrinsic viscosity of each sample was determined at concentrations ranging from 2 to 5 mg mL$^{-1}$, according to equation (1):

$$[\eta] = \lim_{C \to 0} (\eta_{rel} - 1) \cdot \frac{1}{C} \quad (1)$$

where C is the concentration of the solution (g/ml) and $\eta_{rel}$ is the relative viscosity calculated as

$$\eta_{rel} = \frac{\eta}{\eta_0} = \frac{t - \Delta t}{t_0 - \Delta t_0} \quad (2)$$

where η and $\eta_0$ are the viscosity of the solution and of the solvent mixture, respectively, while t is the solution flow time and $t_0$ the solvent mixture flow time in the viscometer.

Five measurements were performed at each concentration for each pCBT sample to reduce the experimental error.

## 2.5. Molecular weight determination

The viscosity-average molecular weight, $M_v$, of the samples was calculated from the intrinsic viscosity [η] values, using the Mark-Houwink equation:

$$[\eta] = K \cdot M_v^{\alpha} \quad (3)$$

where K and α are viscometric parameters which depends on polymer, solvent and temperature. For pCBT, K and α values of $1.17 \cdot 10^{-2}$ mL/g and 0.87, respectively [47, 48].

### 2.6. Characterization

Morphology of nanoflakes was investigated on a high resolution Field Emission Scanning Electron Microscope (FESEM, ZEISS MERLIN 4248) by directly depositing the nanoplatelets on adhesive tape.

Raman spectra were acquired on a Renishaw inVia Reflex (Renishaw PLC, UK) microRaman spectrophotometer equipped with a cooled charge-coupled device camera at excitation wavelength of 514.5 nm with a laser power of 10 mW (spectral resolution and integration time of 3 cm$^{-1}$ and 10 s, respectively). The samples were prepared by drop casting a suspension of GNP in $CHCl_3$ (0.1 mg/mL) on a Si substrate with a 285 nm oxide layer. Spectra were collected on five different particles randomly selected by means of an optical microscope coupled to the instrument.

X-ray Photoelectron Spectroscopy (XPS) was implemented on a VersaProbe5000 Physical Electronics X-ray photoelectron spectrometer equipped with a monochromatic Al K-alpha X-ray source (15 kV voltage, 1486.6 eV energy and 1 mA anode current). Survey scans as well as high resolution spectra were recorded with a 100 μm spot size. Carbon nanoflakes were fixed on adhesive tape and kept under vacuum overnight to remove volatiles. Then, characterization was performed directly on nanoflakes, without any further preparation. Deconvolution of XPS peaks was performed with a Voigt function (Gaussian/Lorentzian = 80/20) after Shirley background subtraction.

Thermogravimetric analysis (TGA) was carried out on a Q500 (TA Instruments, USA). Samples were heated from 50 to 850 °C at the rate of 10 °C min$^{-1}$ in air (gas flow 60 ml min$^{-1}$). The data collected were $T_{max}$ (temperature at maximum rate of weight loss), $T_{onset}$ (the temperature at which the mass lost is 3% of the initial weight) and final residue at 850 °C. TGA on GNP was performed using about 2.5 mg samples.

Differential scanning calorimetry (DSC) was performed on a Q20 (TA Instruments, USA) in the temperature range 50 to 250 °C with a heating rate of 20°C/min. Samples (5 ± 0.5 mg) were first heated to erase the thermal history of the material, then cooled to study the crystallization behavior and, finally, heated again to evaluate melting behavior. Crystallinity was calculated as the ratio between the integrated value for heat of crystallization of the sample and the heat of melting of 100% crystalline PBT, *i.e.* 140 J/g [49]; for nanocomposites, crystallinity degree calculation was performed taking into account the effective polymer fraction in the material, i.e. 95 wt.%.

Rheological properties of pCBT_GNP nanocomposites were evaluated on a strain-controlled rheometer (ARES, TA Instruments, USA) with parallel-plate geometry (25 mm plate diameter). A convection oven, coupled to the instrument was used to control the temperature. Before measurements, dried nanocomposites were pressed at 250 °C into disks with 1 mm thickness and 25 mm diameter. Specimens were further dried at 80 °C in vacuum for 8 h before the measurement to avoid water absorption. Oscillatory frequency sweeps in the linear viscoelastic region ranging from 0.1 to 100 rad/s with a fixed strain (chosen by oscillatory strain sweep tests) were performed in air at 250 °C, to investigate the viscosity of the nanocomposites.

Electrical conductivity (volumetric) was evaluated on 1 mm thick and 25 mm diameter specimens. The apparatus for the measurement is composed by a tension and direct current regulated power supply (PR18-1.2A of Kenwood, Japan), a numeral table multimeter (8845A of Fluke, Everette/USA) equipped with a digital filter for noise reduction, a palm-sized multimeter (87V of Fluke, Everette/USA) and two homemade brass electrodes: a cylinder (18.5 mm diameter, 55 mm height) and a square plate (100 mm side, 3mm thickness). Every electrode has a hole for the connection and a wire equipped with a 4mm banana plug. Power supply is time to time regulated in current or in voltage to measure accurately, with both the multimeters, limiting the power dissipated on the specimen. Electrical conductivity was calculated with the following formula

$$\sigma = \frac{1}{\rho} = \frac{I}{V} \cdot \frac{l}{S} \; [S/m] \tag{4}$$

where $\rho$ is the electrical resistivity, $l$ and $S$ are specimen thickness and surface, respectively; $I$ is the electric current and $V$ the voltage, both read by the measurement apparatus.

Isotropic thermal conductivity tests were executed on a TPS 2500S (Hot Disk AB, Sweden). Before the measurement, specimens (4 mm thickness and 15 mm diameter) were stored in a constant climate chamber (Binder KBF 240, Germany) at $23.0 \pm 0.1$ °C and $50.0 \pm 0.1$ %R.H. for at least 48 h. Measuring temperature ($23.00 \pm 0.01$ °C) was controlled by a silicon oil bath (Haake A40, Thermo Scientific Inc., USA) equipped with a temperature controller (Haake AC200, Thermo Scientific Inc., USA).

## 3. Results and discussion

### 3.1. Nanoflakes characterization

In this work, the effect of processing parameters on thermal conductivity of pCBT/GNP nanocomposites was studied. In a previously published paper [30], we reported that the quality of graphene-related materials (i.e. defectiveness and aspect ratio) is crucial to obtain highly thermally conductive polymer nanocomposites; indeed, morphology and structural defects of GNP were thoroughly characterized and the results will be reported and discussed below.

The morphology of the as received GNP, showed in Figure 1a, reveals few nanometer thick wrinkled layers, organized in accordion-like structures of some mm length and ~ 200 μm lateral size. However, high magnification micrographs show separated nanoflakes (Figure 1b) on the scale of some tens of micrometers and thickness estimated in few nanometers. It is worth considering that a similar expanded structure is typical of the thermal expansion process, as observed for thermally reduced graphene oxide [30].

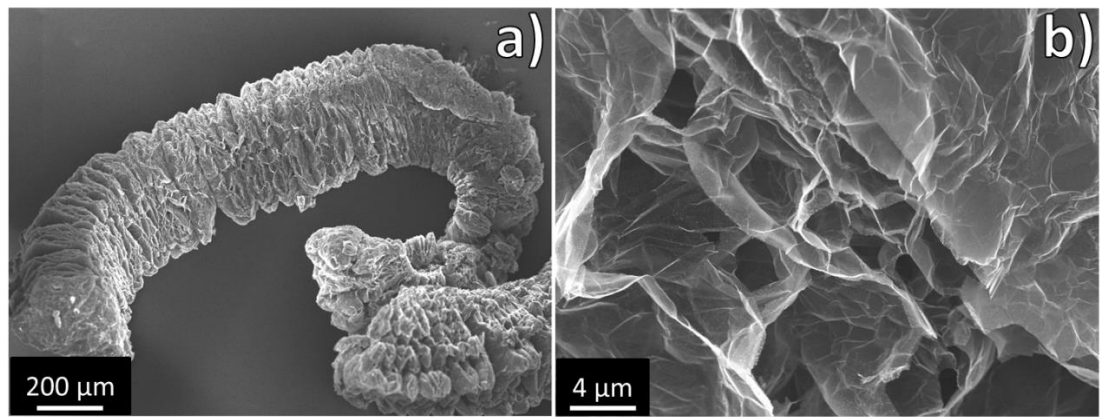

**Figure 1. FESEM micrographs on GNP: (a) low and (b) high magnification**

Chemical composition and defectiveness of nanoflakes were evaluated by means of XPS, Raman spectroscopy and TGA. The oxygen content of GNP, calculated from the integration of survey scan peaks of XPS data, was ~ 5.0 at.%, thus indicating a C/O ratio of about 19/1. A deeper insight on the functional groups was performed by deconvolution of $C_{1s}$ (B.E. ≈ 285 eV, Figure 2a) and $O_{1s}$ (B.E. ≈ 530 eV, Figure 2b) peaks, collected by narrow scans. $C_{1s}$ spectra (Figure 2a) show an intense peak located at ~ 284.2 eV, assigned to $sp^2$ C-C carbon, and a long tail which was deconvolved with five peaks centered at: ~ 284.7 eV ($sp^3$ C-C carbon), ~ 285.6 eV (C-OH, C-O-C), ~ 286.7 eV (C=O), ~ 287.7 eV (HO-C=O) and ~ 290.8 eV (π-π* shake-up of the aromatic carbon) [50-52]. It is worth noting that the relatively intense peak located at ~ 290.8 eV and a narrow $sp^2$ C-C peak (FWHM ≈ 0.69 eV) are typically related to the presence of a high aromaticity degree in the graphitic structure [27, 52]. Deconvolution of $O_{1s}$ signal is reported in Figure 2b: a reliable fitting into two peaks was obtained, thus indicating the coexistence of single-bonded (~ 533.0 eV, C-OH/C-O-C) and double-bonded oxygen (~ 531.4 eV, C=O/O=C-OH), in agreement with deconvolution of $C_{1s}$ signal.

Representative Raman spectrum for GNP, normalized with respect to the G peak (~ 1573 $cm^{-1}$), is displayed in Figure 2c. First-order Raman spectrum shows a tiny D band at ~ 1358 $cm^{-1}$ and a strong and narrow G band. This results in a low $I_D/I_G$ ratio ($I_D/I_G$ ≈ 0.06, calculated from the intensities of fitting peaks), thus indicating low defectiveness of GNP [27], in agreement with XPS results. The second-order spectrum shows the presence of an intense band located at about 2710 $cm^{-1}$ (G' band), which is deconvolved into two main peaks located at ~ 2682 $cm^{-1}$ ($G'_1$) and ~ 2715 $cm^{-1}$ ($G'_2$), respectively. Both bands are characteristic of graphene-related materials constituted by more than five graphene layers [53].

The thermal stability of GNP, and hence a further qualitative evaluation of its defectiveness, was evaluated by TGA in air (Figure 2d); in fact, it was reported that size and defectiveness, of graphene-related materials, affect the onset of decomposition temperature [54]. The thermogram of the GNP used in this study exhibits two degradation steps: in the first, weight loss of about 6 wt.% occurred between ~ 450°C and 600°C, which could be related to smaller and highly defective nanoparticles; whereas in the second step a further 80 wt.% loss is verified between ~ 600°C and 850°C, with the maximum of mass loss rate centered at ~ 762 °C and about 11 wt.% residue at 850°C, thus indicating an high content of large and low defective nanoflakes, according to the work of Shtein et al [54]. This is crucial for the obtainment of highly thermally conductive polymer nanocomposites.

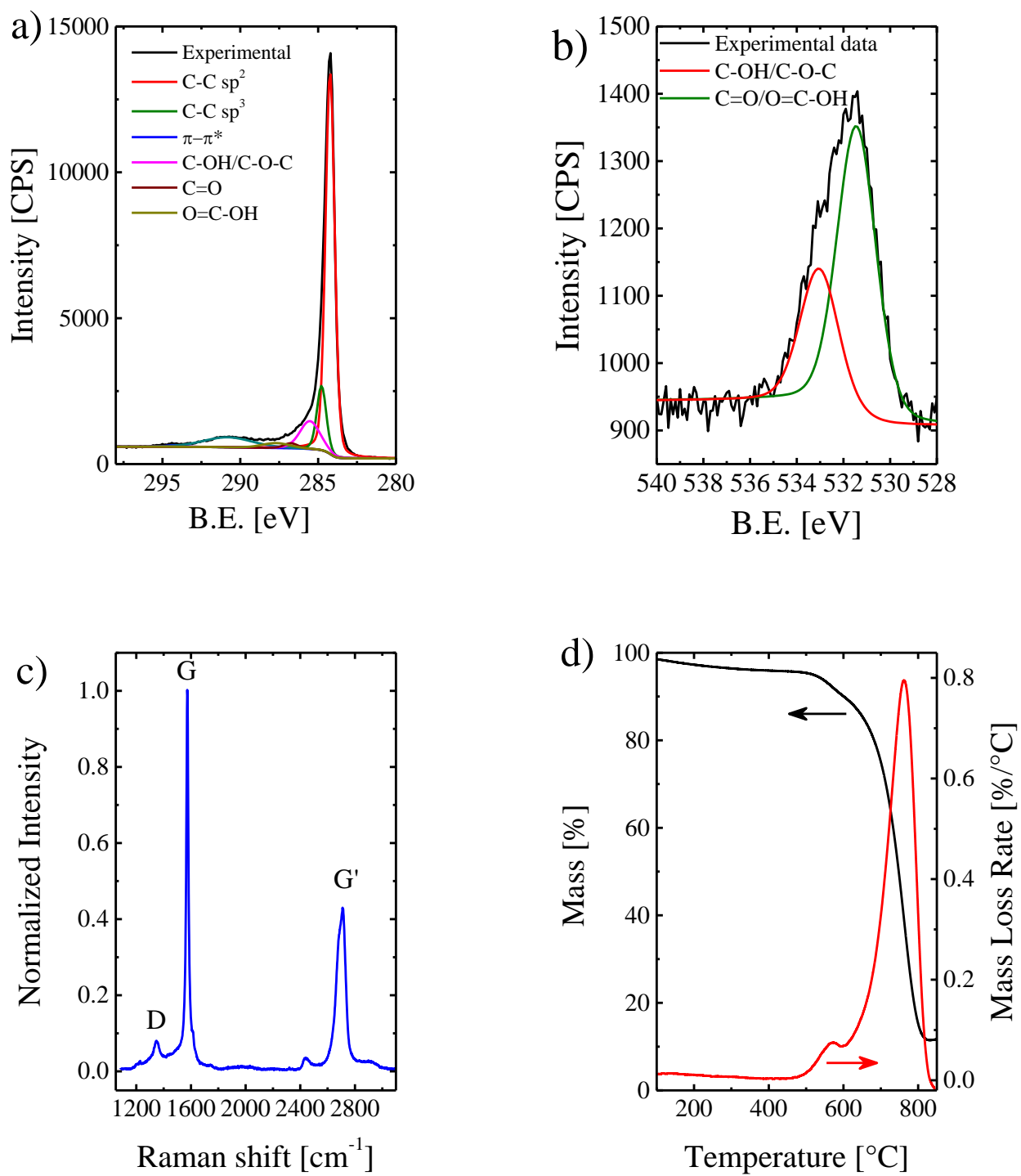

**Figure 2.** Evaluation of GNP properties: (a) C1s and (b) O1s XPS curves and respective deconvolutions, (c) representative Raman spectrum normalized with respect to G peak and (d) thermogravimetric analysis results.

### 3.2. Nanocomposites characterization

The polymerization of CBT into pCBT, and the effect of the processing conditions on melting/crystallization behavior of the different nanocomposites, were monitored by differential scanning calorimetry (Figure S1 and Table S1). Heating curves demonstrate that no residual traces of the characteristic melting peaks of the oligomer were observed for all the nanocomposites. Furthermore, in heating scans, the presence of only one endothermic peak for pCBT/GNP nanocomposites indicates the formation of stable crystals during cooling. Instead, in pure pCBT the presence of an additional endothermic peak at lower temperatures was related to melting/recrystallization of imperfect crystals [55, 56]. The presence of carbon nanoflakes usually affects the crystallization behavior of the polymer matrix [30]: in fact, while pure pCBT has a crystallization temperature of 192.2°C, values ranging between 195.2°C and 204.6°C were measured for pCBT + GNP nanocomposites, the highest for pCBT_GNP/240/10/100, thus indicating a pronounced nucleating effect of GNP. Crystallinity degree (Table S1) of pure pCBT was estimated equal to 49.2%, while for nanocomposites values ranging between 45.9% and 53.6% were calculated. The comparison between the different materials reveals that none of the three processing parameters (temperature, time and shear rate) or their combination have a clear effect on neither crystallinity nor crystallization peak temperature.

The average viscosimetric molecular weight [57] of pCBT samples was determined, after GNP extraction, from the intrinsic viscosity of pCBT solutions and results are summarized in Table 2. The value of the $M_v$ calculated for pCBT was 40500 g/mol, achieving a sufficient polymerization degree of pCBT using the extrusion process. The presence of graphene nanoparticles affects the ring-opening polymerization of CBT, with a general reduction in the molecular weight of the polymer matrix, in agreement with Fabbri et al. [44]. Indeed, the average molecular weight of pCBT including 5 wt.% of GNP decreased in more than 40% respect to the value of the neat pCBT, for all nanocomposites. None of the processing parameters or their combination exhibits major effects on the final molecular weight of the neat polymer, thus indicating that $M_v$ is mainly affected by the presence of nanoparticles. However, it is worth observing that the combination of high processing time and low temperature leads to the obtainment of the highest molecular weight (Figure S2) suggesting that low polymerization temperature is beneficial to reduce chain scission during mixing.

**Table 2. Viscosity values extrapolated for pCBT and pCBT_GNP nanocomposites**

| Nanocomposite | $M_v$ [kg/mol] |
|---|---|
| **pCBT** | 40.5 ± 0.5 |
| **pCBT_GNP/240/5/50** | 24.8 ± 0.1 |
| **pCBT_GNP/240/10/50** | 27.1 ± 0.1 |
| **pCBT_GNP/240/5/100** | 28.7 ± 0.1 |
| **pCBT_GNP/240/10/100** | 28.5 ± 0.3 |
| **pCBT_GNP/260/5/50** | 25.5 ± 0.1 |
| **pCBT_GNP/260/10/50** | 25.8 ± 0.1 |
| **pCBT_GNP/260/5/100** | 22.9 ± 0.2 |
| **pCBT_GNP/260/10/100** | 27.6 ± 0.3 |

The study of GNP dispersion and distribution, in pCBT, was performed by means of linear viscoelasticity in the molten state; indeed, complex viscosity, $\eta^*$, and elastic modulus, G', are well-known to be related to nanoparticle amount and organization in a polymer matrix [58, 59]. G' and $\eta^*$ plots for pCBT and its nanocomposites, obtained from dynamic frequency tests, are reported in Figure 3 as a function of deformation frequency. pCBT exhibits the typical behavior of pure polymers in linear regime, with viscosity approximately independent on the frequency and modulus decreasing when frequency decreases; it is worth noting that, owing to the instrument limits, it was not possible to evaluate G' (for pure pCBT) at frequencies below $\omega \approx 6$ rad/s. The inclusion of GNP leads to higher G' and $\eta^*$ in the whole frequency range: for all the nanocomposites, a strong shear thinning effect (decrease of the viscosity as frequency increases, extending over two decades in the explored frequency range) and a weak G' dependence on the frequency were observed, thus indicating the formation of a percolated structure constituted by graphite nanoplatelets [30, 59, 60]. Furthermore, the formation of the solid like network in pCBT nanocomposites is evidenced by a predominance of the elastic response in the whole frequency range, while in pure pCBT viscoelastic properties are mainly dominated by the viscous response (Figure S3). To further investigate the effect of the different processing conditions on the viscoelastic properties of pCBT_GNP nanocomposites, G' and $\eta^*$ values at low frequencies ($\omega \approx 1$ rad/s) were compared (Table S2). Results show that viscoelastic properties of pCBT nanocomposites were weakly affected by the extrusion temperature; in fact, for nanocomposites prepared with the same processing time and shear rate, slightly higher modulus and viscosity values were measured when lower temperatures were used for nanocomposite preparation (i.e., G' and η* values of ~ 3.1 · 10³ Pa and ~ 3.7 · 10³ Pa·s, respectively, were measured for pCBT_GNP/240/5/50, whereas pCBT_GNP/260/5/50 showed G' ≈ 1.1 · 10³ Pa and η* ≈ 1.5 · 10³ Pa·s), suggesting a denser percolation network obtained at the lower processing temperature. On the other hand, no significant trends on both viscosity and elastic modulus were observed varying time or shear rate.

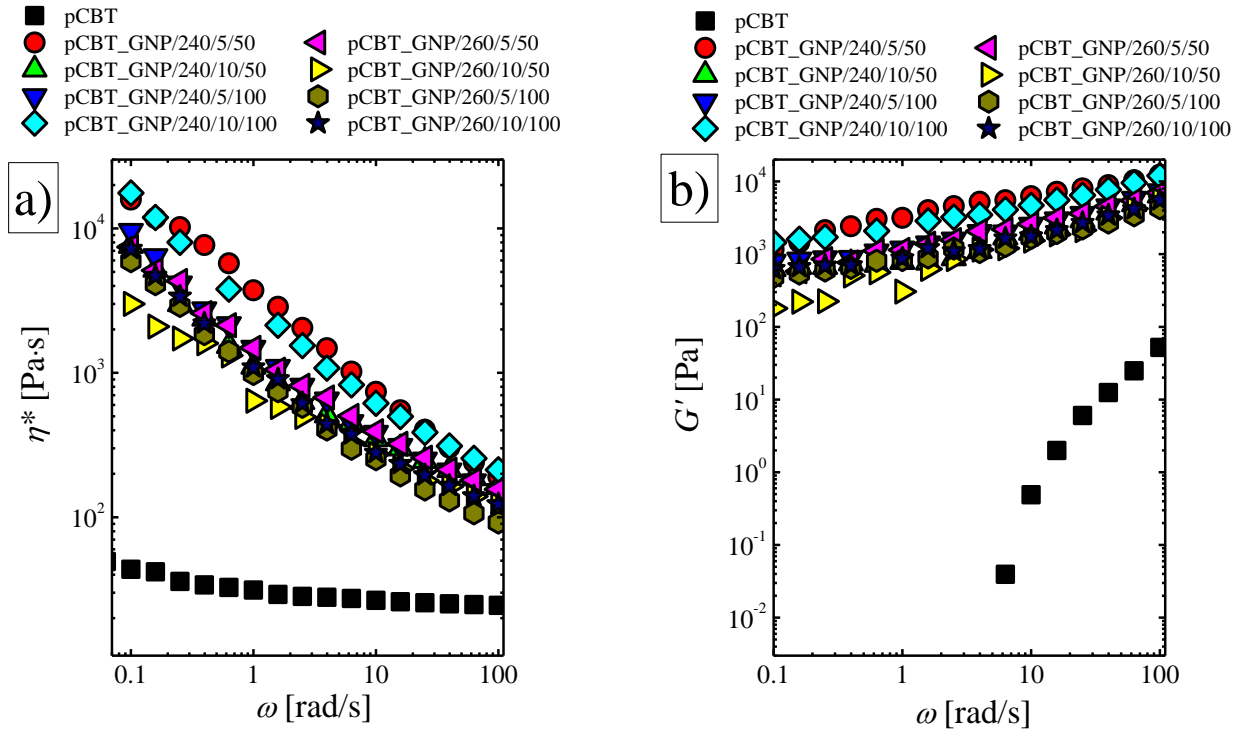

**Figure 3. Dynamic frequency sweep test at 250°C for pCBT and pCBT + GNP nanocomposites. (a) η* and (b) G' as a function of the angular frequency**

The organization of conductive particles in a solid-like network is crucial for the preparation of electrically conductive materials [30, 61]: in particular, the higher the density of the percolation network, the higher the electrical conductivity. Pure pCBT displays an electrical conductivity in the range of $10^{-13}$ S/m [43], which is typical for pure polymers, whereas the addition of GNP leads to a sharp increase (see Figure 4 and Table 3) with values ranging between 2.75 · $10^{-5}$ and 5.89 · $10^{-3}$ S/m for pCBT_GNP/260/10/100 and pCBT_GNP/240/5/50, respectively, thus indicating that all the nanocomposites are well above the electrical percolation threshold. It is clearly observable that the highest and the lowest electrical conductivity values were measured using the milder (low time, low shear rate) and the severer (high time, high shear rate) processing conditions, respectively. Temperature is also affecting the conductivity values, generally leading to higher electrical performance with lower processing temperature despite the opposite effect was observed in pCBT_GNP/5/10. These suggest a strong effect of the harsher processing conditions, with a likely reduction of the GNP aspect ratio during extrusion for long time and/or high shear rates and/or high temperatures [34, 62]. In fact, nanoparticle aspect ratio was reported to affect electrical conductivities of polymer nanocomposites, with higher value obtained when high aspect ratio nanoparticles were used [63, 64].

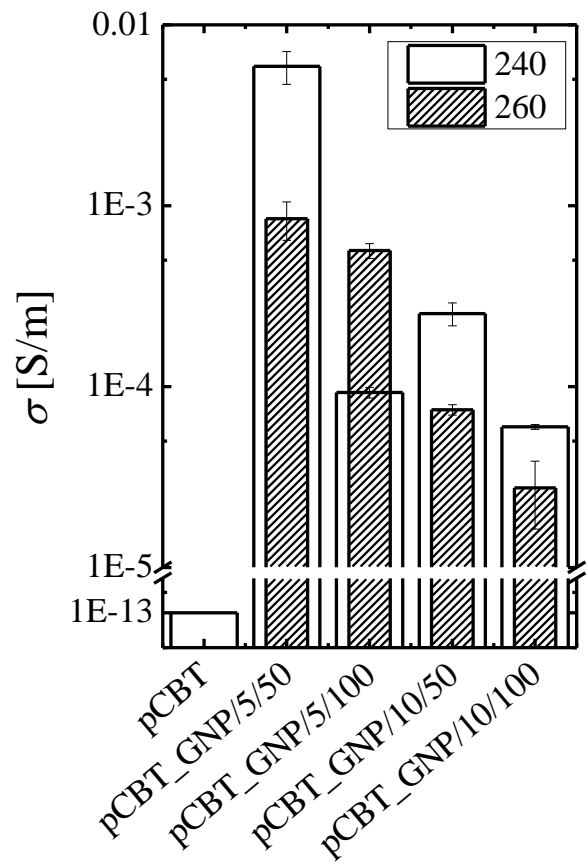

**Figure 4. Electrical conductivity of pCBT + GNP as function of the different extrusion parameters**

Bulk thermal conductivity results for pCBT and pCBT_GNP nanocomposites as a function of processing parameters are reported in Figure 5 and Table 3. Pure pCBT exhibits a thermal conductivity of ~ 0.24 W/(m·K) which is consistent with results reported in literature for semi-crystalline polymers [1]. The addition of GNP leads to an improvement in the conductivity properties with values ranging between 0.72 and 0.98 W/(m·K) for pCBT_GNP/260/10/100 and pCBT_GNP/240/5/50, respectively, thus indicating an increase between 3-fold and 4-fold, with respect to pure polymer. Thermal conductivity results, as a function of the different parameters, exhibit a trend similar to that observed for electrical conductivities (Figure 4), with the highest conductivity values obtained combining short time and low shear rate (see also Figure S5). Indeed, the reduction in thermal conductivity between the nanocomposite obtained in the mildest conditions (240/5/50) *vs.* the one prepared in the harshest conditions (260/10/100) is as much as 26%. This fact is most likely related to the reduction in nanoflake aspect ratio, which was reported to have a detrimental effect on the improvement of nanocomposites thermal conductivity [24].

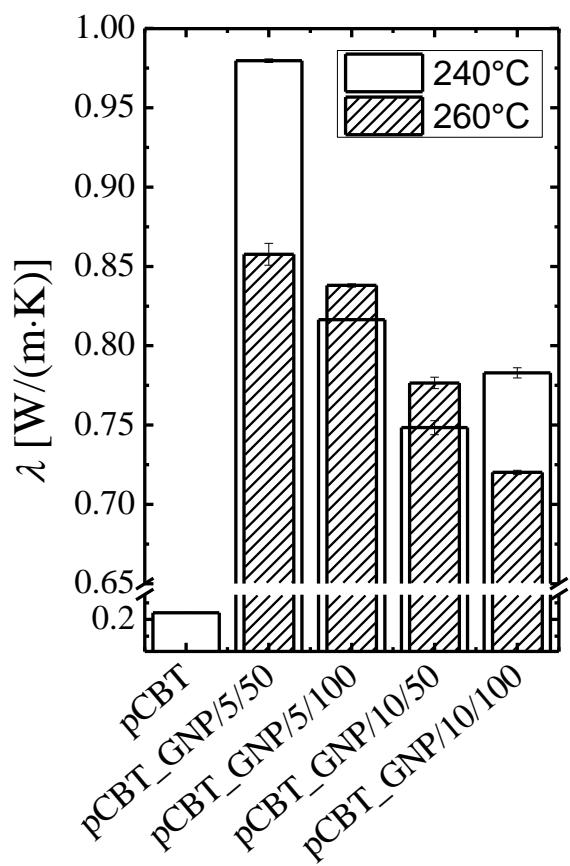

**Figure 5. Thermal conductivity of pCBT and pCBT + GNP as function of the different extrusion parameters**

**Table 3. Electrical and thermal conductivity data for pCBT nanocomposites as a function of processing parameters**

| Electrical conductivity [S/m] ($\sigma$) |
| --- |
| Thermal Conductivity [W/(m·K)] ($\lambda$) |

| Temperature | 240 °C | 260 °C |
| --- | --- | --- |
| pCBT_GNP/5/50 | σ = (5.9 ± 0.1) E-3 | σ = (8.5 ± 2.0) E-4 |
| | λ = 0.980 ± 0.001 | λ = 0.858 ± 0.007 |
| pCBT_GNP/5/100 | σ = (9.3 ± 0.6) E-5 | σ = (5.6 ± 0.5) E-4 |
| | λ = 0.816 ± 0.001 | λ = 0.838 ± 0.001 |
| pCBT_GNP/10/50 | σ = (2.5 ± 0.4) E-4 | σ = (7.4 ± 0.5) E-5 |
| | λ = 0.748 ± 0.004 | λ = 0.777 ± 0.004 |
| pCBT_GNP/10/100 | σ = (6.0 ± 0.2) E-5 | σ = (2.8 ± 1.1) E-5 |
| | λ = 0.783 ± 0.003 | λ = 0.720 ± 0.001 |

Beside the clear effect of mixing time and shear rate, for both electrical and thermal conductivity, the role of processing temperature appears to be complex as in some cases a decrease of conductivity was observed when increasing temperature, whereas the opposite trend was obtained. To further analyse the results obtained for the four main properties addressed (molecular weight, melt viscosity, electrical and thermal conductivity) upon the processing parameters, average values were calculated for properties of the 4 different formulations prepared with one processing parameter as a constant. This averaging was repeated for all the different properties and parameters, leading to average values suitable to compare performances obtained at the low and high setting points for each of the parameter (temperature, mixing time and shear rate). Results of this analysis are reported in Figure 6, in which averaged values calculated as above for low setting points, of the different parameters, were normalized to 1 and averaged values for high setting points were scaled accordingly. From this analysis it is clear that electrical conductivity is the property being affected the most by melt processing conditions: comparing the average value of the electrical conductivities at the lowest and highest setting points, systematically higher electrical conductivities were measured with the use of the lowest setting points, as increase by factors 4, 10 and 19 were obtained for temperature, time and shear rate, respectively, as compared with their highest setting points counterparts.

Beside the effect on electrical conductivity, the increase in mixing time decreases percolation density in the melt, indirectly evaluated from melt viscosity, as well as thermal conductivity, whereas a slight increase in molecular weight was obtained for longer processing time. The screw rotation rate has a larger effect in reducing both melt viscosity and thermal conductivity, with a slight increase of molecular weight. Finally, temperature increase was not found to be beneficial for any of the properties addressed, possibly due to side reactions or chain scission during melt mixing.

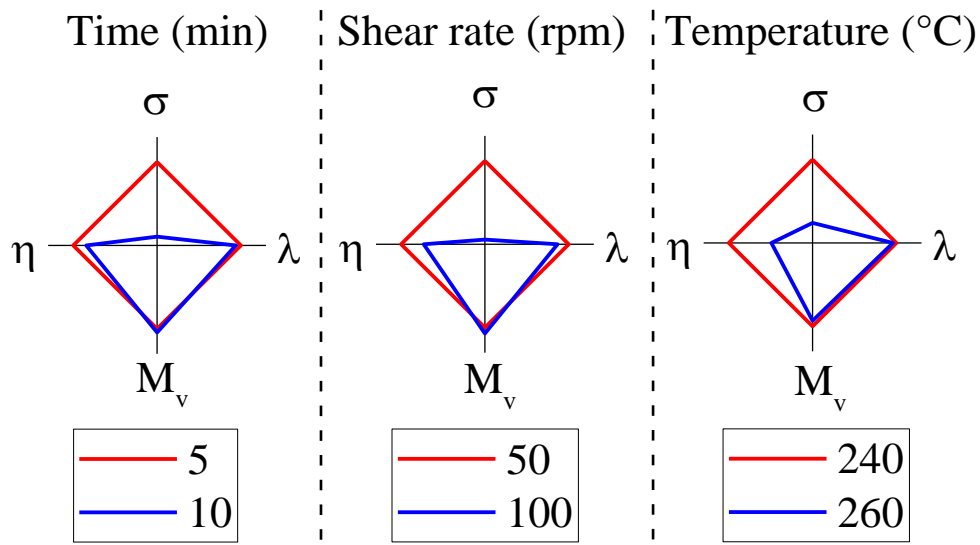

**Figure 6. Effect of time, shear rate and temperature on electrical conductivity, thermal conductivity, molecular weight and viscosity of pCBT + GNP nanocomposites. Each of the axis report averaged and normalized values for comparison between performance of higher vs lower setting points. Scale on electrical conductivity is logarithmic.**

Based on the results here showed it appears clear that the use of mild processing conditions is essential to maximize electrical and thermal properties of pCBT_GNP nanocomposites. Despite this work was not aimed to a full optimization of processing conditions, which would require both enlarging processing windows and considering additional parameters (e.g. screw profile), the results obtained should be taken into account when designing up-scaling of nanocomposites production onto industrial scale equipment, which is clearly beyond the scope of this paper.

Beside the effect of the addressed processing parameters, electrical and thermal conductivities are typically strongly affected by the GNP content, increasing the amount of nanoparticle content is clearly expected to improve especially thermal conductivity. However, higher amount of GNP is directly related to a significant increase in viscosity, which may also restrict the possible processing window.

4. Conclusions

In this work, the effect of different processing paremeters on the properties of poly (butylene terephthalate) nanocomposites prepared via ring-opening polymerization of CBT in presence of graphite nanoplatelets were addressed. In particular, the present paper is focused on the effects of processing temperature, mixing time and shear rate on polymer molecular weight, nanoparticle dispersion as well as electrical and thermal conductivity of pCBT/GNP nanocomposites.

Average viscosimetric molecular weight of pCBT was found to be significantly affected by the presence of nanoflakes, with a general reduction in the molecular weight, compared to pure pCBT, in the range of 40%, regardless of the processing parameters used for compounding. Despite the limited molecular weight obtained, a satisfactory dispersion and distribution of GNP was observed, with the formation of a dense percolating networks, evidenced by the study of linear viscoelasticity in the molten state.

Electrical and thermal conductivity results showed similar trends with the highest conductivity values ($\sigma \approx 6 \cdot 10^{-3}$ S/m and $\lambda \approx 1.0$ W/(m·K), respectively) obtained combining short time, low temperature and low shear rate, whereas the lowest values were obtained ($\sigma \approx 3 \cdot 10^{-5}$ S/m and $\lambda \approx 0.7$ W/(m·K), respectively) setting the three parameters at the higher level (harsher processing conditions). These observations were related to the reduction of nanoflake aspect ratio upon ring-opening polymerization for longer time and greater shear rates. These results evidences the need for careful optimization of processing parameters during preparation of polymer nanocomposites containing graphene related materials, a field in which, to the best of the authors' knowledge, no other work was reported so far on the systematic study of processing conditions and correlation with the properties of materials.

**Authors contributions**

A. Fina conceived the experiments and coordinated the project, G.Gavoci, M.M. Bernal and S. Colonna carried out the preparation of nanocomposites and most of the characterizations reported in this paper. J. Gomez prepared graphene nanoplatelets. C.Novara performed Raman spectroscopy measurements. G.

Saracco contributed to the discussion of the results. Manuscript was written by S. Colonna, M.M. Bernal and A. Fina.

**Acknowledgements**

The research leading to these results has received funding from the European Union Seventh Framework Programme under grant agreement n°604391 Graphene Flagship. This work has received funding from the European Research Council (ERC) under the European Union's Horizon 2020 research and innovation programme grant agreement 639495 — INTHERM — ERC-2014-STG. Funding from Graphene@PoliTo initiative of the Politecnico di Torino is also acknowledged.

The authors gratefully acknowledge Mauro Raimondo for FESEM observations, Salvatore Guastella for XPS analyses, Fabrizio Giorgis for Raman Spectroscopy as well as Fausto Franchini for electrical conductivity measurements. Orietta Monticelli from University of Genova is also gratefully acknowledged for useful discussions on the ring-opening polymerization process.

# SUPPLEMENTARY MATERIAL

# Effect of processing conditions on the thermal and electrical conductivity of poly (butylene terephthalate) nanocomposites prepared via ring-opening polymerization

S. Colonna[a], M.M. Bernal[a], G. Gavoci[a], J. Gomez[b], C. Novara[c], G. Saracco[d], A. Fina[a,*]

[a]*Dipartimento di Scienza Applicata e Tecnologia, Politecnico di Torino, 15121 Alessandria, Italy*
[b]*AVANZARE Innovacion Tecnologica S.L., 26370 Navarrete (La Rioja), Spain*
[c]*Dipartimento di Scienza Applicata e Tecnologia, Politecnico di Torino, 10129 Torino, Italy*
[d]*Istituto Italiano di Tecnologia, Centre for Sustainable Futures CSF@PoliTo, 10129 Torino, Italy*
*\*Corresponding author: alberto.fina@polito.it*

**Differential scanning calorimetry**

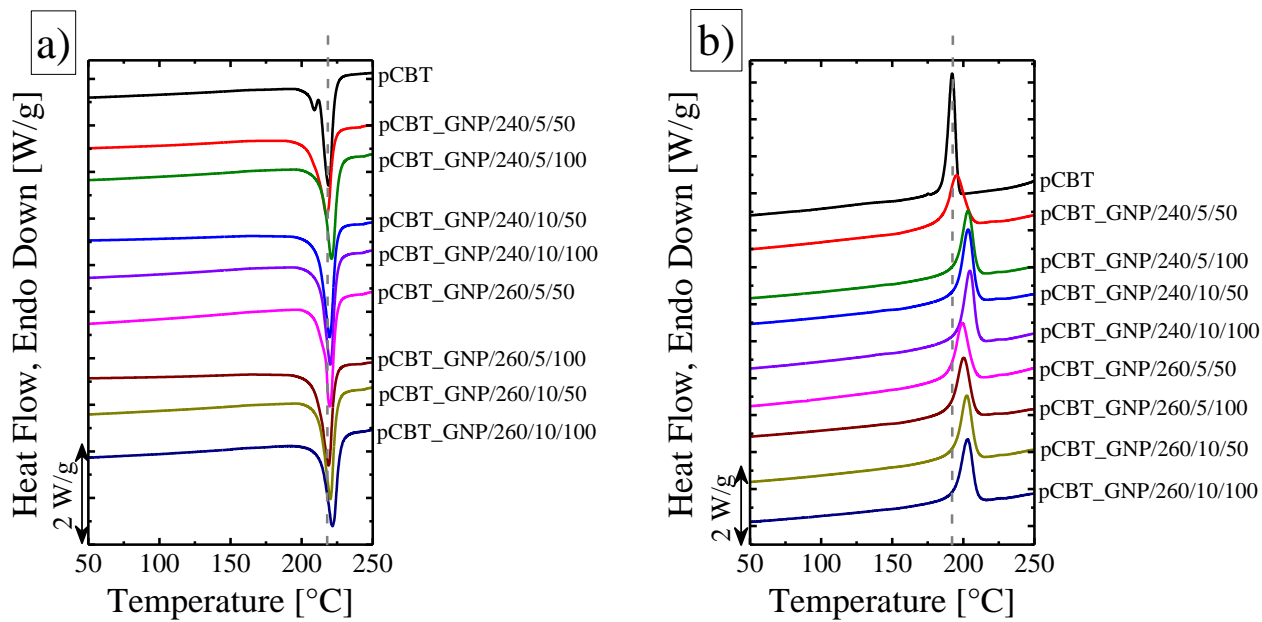

**Figure S1**. DSC curves for pCBT and pCBT_GNP nanocomposites. a) Heating and b) cooling, both at 20°C/min

**Table S1**. Melting temperature ($T_m$), crystallization temperature ($T_c$) and crystallinity ($X_c$) for pCBT and pCBT_GNP nanocomposites

| Nanocomposite | $T_m$ [°C] | $T_c$ [°C] | $X_c$ [%] |
|---|---|---|---|
| **pCBT** | 219.1 | 192.2 | 49.2 |
| **pCBT_GNP/240/5/50** | 217.5 | 195.2 | 48.4 |
| **pCBT_GNP/240/10/50** | 219.8 | 203.4 | 53.6 |
| **pCBT_GNP/240/5/100** | 221.1 | 203.4 | 47.0 |
| **pCBT_GNP/240/10/100** | 220.2 | 204.6 | 50.5 |
| **pCBT_GNP/260/5/50** | 219.9 | 199.4 | 45.9 |
| **pCBT_GNP/260/10/50** | 220.3 | 202.5 | 46.0 |
| **pCBT_GNP/260/5/100** | 219.1 | 200.3 | 49.1 |
| **pCBT_GNP/260/10/100** | 221.8 | 203.3 | 46.7 |

**Molecular weight determination**

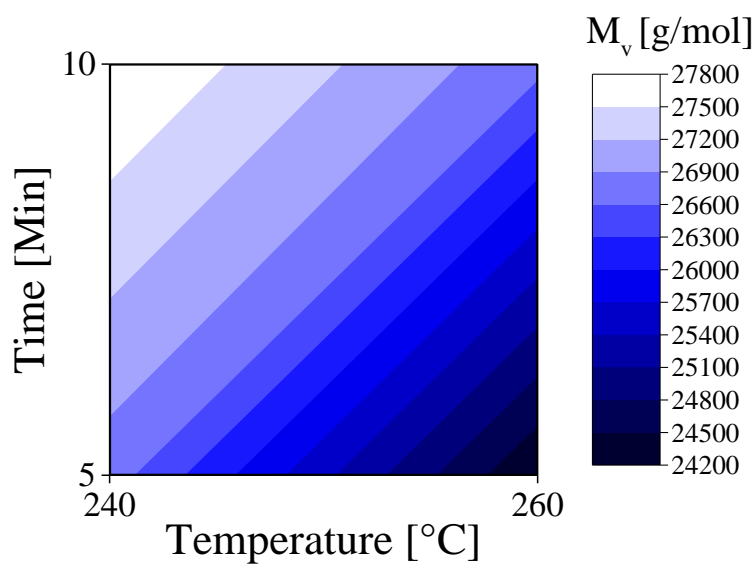

**Figure S2**. Contour plots for the molecular weight results of pCBT_GNP nanocomposites as a function of temperature and time. Values at each corner were obtained by averaging the two values at 50 and 100 rpm in the selected *x* temperature and *y* time. Gradient fill color is a guideline for the eyes

**Rheological analysis**

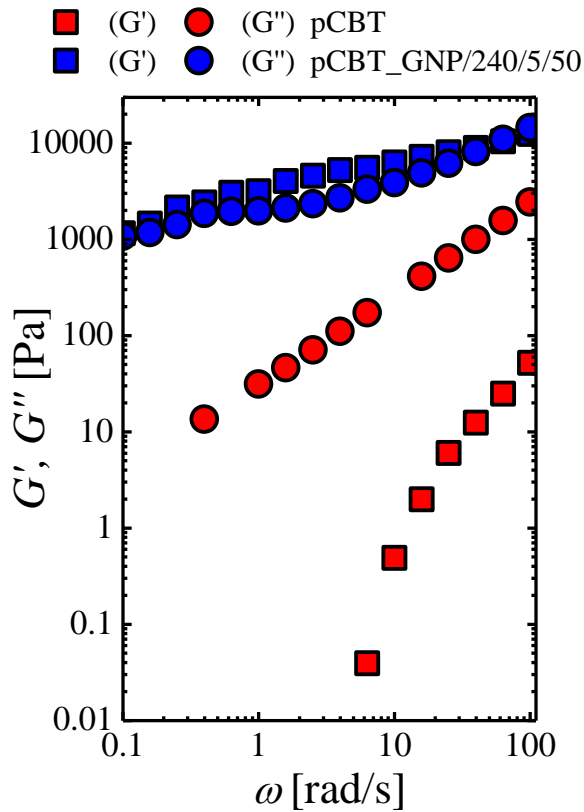

**Figure S3**. G' and G" vs. ω for pCBT and pCBT_GNP/240/5/50

**Table S1**. Elastic modulus and viscosity (both measured at ω ≈ 1 rad/s) for pCBT and pCBT_GNP nanocomposites

| Nanocomposite | G' (ω ≈ 1 rad/s) [Pa] | η* (ω ≈ 1 rad/s) [Pa·s] |
|---|---|---|
| **pCBT** | - | 30 |
| **pCBT_GNP/240/5/50** | 3164 | 3732 |
| **pCBT_GNP/240/10/50** | 783 | 1103 |
| **pCBT_GNP/240/5/100** | 1191 | 1435 |
| **pCBT_GNP/240/10/100** | 2422 | 2578 |
| **pCBT_GNP/260/5/50** | 1141 | 1498 |
| pCBT_GNP/260/10/50 | 305 | 812 |
| pCBT_GNP/260/5/100 | 831 | 963 |
| pCBT_GNP/260/10/100 | 874 | 1071 |

**Electrical conductivity**

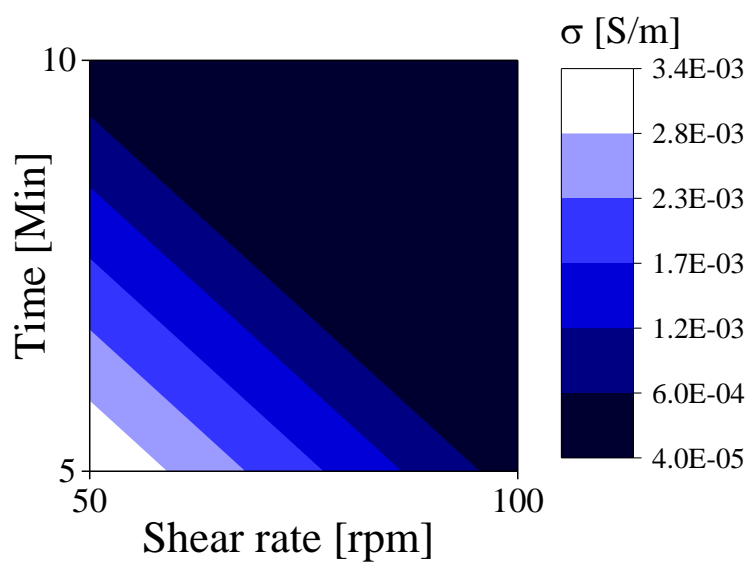

**Figure S4**. Contour plots for the electrical conductivity of pCBT_GNP nanocomposites as a function of time and shear rate. Values at each corner were obtained by averaging the two values at 240 and 260°C in the selected *x* shear rate and *y* time. Gradient fill color is a guideline for the eyes

**Thermal conductivity**

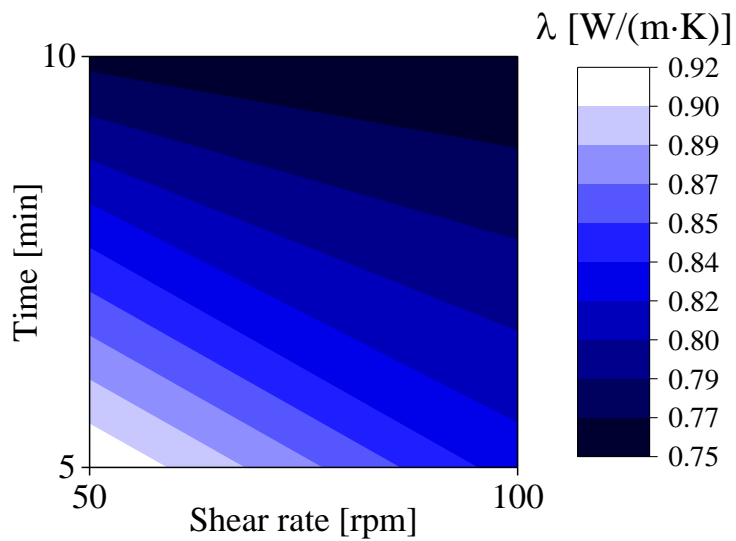

**Figure S5**. Contour plots for the thermal conductivity of pCBT_GNP nanocomposites as a function of time and shear rate. Values at each corner were obtained by averaging the two values at 240 and 260°C in the selected *x* shear rate and *y* time. Gradient fill color is a guideline for the eyes